\documentclass[sigconf]{acmart}
\usepackage{csquotes}

\AtBeginDocument{%
  \providecommand\BibTeX{{%
    \normalfont B\kern-0.5em{\scshape i\kern-0.25em b}\kern-0.8em\TeX}}}

\copyrightyear{2020} 
\acmYear{2020} 
\setcopyright{acmlicensed}
\acmConference[SIGCSE '20]{The 51st ACM Technical Symposium on Computer Science Education}{March 11--14, 2020}{Portland, OR, USA}
\acmBooktitle{The 51st ACM Technical Symposium on Computer Science Education (SIGCSE '20), March 11--14, 2020, Portland, OR, USA}
\acmPrice{15.00 \textbf{Pre-print version shared by authors.}}
\acmDOI{10.1145/3328778.3366921}
\acmISBN{978-1-4503-6793-6/20/03}

\begin{document}

\title{Effects of Human vs. Automatic Feedback on Students' Understanding of AI Concepts and Programming Style}

\author{Abe Leite}
\affiliation{
  \institution{Cognitive Science Program\\Indiana University, Bloomington}
 }
\email{abrleite@indiana.edu}

\author{Sa\'ul A. Blanco}
\affiliation{
  \institution{Department of Computer Science\\Indiana University, Bloomington}
}
\email{sblancor@indiana.edu}

\begin{abstract}
  The use of automatic grading tools has become nearly ubiquitous in large undergraduate programming courses, and recent work has focused on improving the quality of automatically generated feedback. However, there is a relative lack of data directly comparing student outcomes when receiving computer-generated feedback and human-written feedback. This paper addresses this gap by splitting one 90-student class into two feedback groups and analyzing differences in the two cohorts' performance. The class is an intro to AI with programming HW assignments. One group of students received detailed computer-generated feedback on their programming assignments describing which parts of the algorithms' logic was missing; the other group additionally received human-written feedback describing how their programs' syntax relates to issues with their logic, and qualitative (style) recommendations for improving their code. Results on quizzes and exam questions suggest that human feedback helps students obtain a better conceptual understanding, but analyses found no difference between the groups' ability to collaborate on the final project. The course grade distribution revealed that students who received human-written feedback performed better overall; this effect was the most pronounced in the middle two quartiles of each group. These results suggest that feedback about the syntax-logic relation may be a primary mechanism by which human feedback improves student outcomes.
\end{abstract}



\maketitle

\section{Introduction}

The use of automatic grading tools has become nearly ubiquitous in large undergraduate programming courses to accommodate for increased enrollment in traditional CS programs as well as online courses. Several authors have addressed automatic grading platforms~\cite{BF16, EPQ08,Geigleetal,HW69,IAK10,Milojicic11,SA14}. Many of the systems used so far have been focused on binary criteria: Does the program run?, does the program produce the right output for problem 1?, does the program produce the right output for problem 2?, etc. These systems offer limited feedback to the student. Recently there has been some effort to provide better feedback using automatic grading~\cite{Choudhuryetal16,Gutierrezetal12,Haldemanetal18,Pariharetal07}. In particular, in~\cite{Haldemanetal18}, the authors utilize existing autograding systems and designed knowledge maps to classify common errors to automatically generate hints meant to address said errors.

However, the current body of work is focused on introductory computer science classes. Our work here focuses on a 300-level class where the programming assignments are a bit more involved, as well as the concepts students are implementing. More than mastery of a programming language, we were interested in assessing students' understanding of the algorithmic concepts covered in class as well as programming style according to ~\cite{PEP8}. Moreover, we found that current autograding systems did not provide ways to offer feedback at the level of granularity we desired to provide to our students. Given that our class is focused on implementing AI algorithms, we hoped to provide students with a checklist of the steps and concepts necessary for a working implementation of each algorithm, and then evaluate them using this checklist to assist them in improving their code. Our teaching staff developed an in-house grading tool for intermediate classes that we plan to share to the larger CS community.

Additionally, much of the current literature assesses grading approaches according to the similarity of grades and feedback to human feedback~\cite{BF16, Geigleetal,Haldemanetal18,SA14}, rather than by student outcomes. Given the extent to which automatic grading has become commonplace, it is natural to ask what (if anything) is lost in the real world by replacing human grading with automatic feedback. A primary aim of our current research is to answer this question.

So the two main contributions of our work are: (i) an open-source grading tool that can be used by any CS educator who would like to provide more granularity in the automatic feedback generated, and (ii) a direct comparison of students' understanding of the algorithmic concepts from lecture as well as programming style, when graded by a very informative automatic grading system and when additional feedback is provided by human graders (TA). We discuss our findings in detail in Section~\ref{s:discussion}.

\subsection{Description of the Class}
This is a 300-level class that serves as an introduction to artificial intelligence, covering topics such as logic, constraint satisfaction problems, search algorithms, games, decision trees, and neural networks. The class is taught every semester and there are usually around 80 students that finish the class. The only official prerequisite is an introductory programming class, though in practice several students have also taken a discrete math class, a second programming class, or even a data structures class. Students work on individual assignments that contain both a programming and a theoretical component. The programming component is implemented in Python and is designed to be a mini-project, such as building a Sudoku solver or implementing an AI to play Connect Four. Each mini-project includes significant scaffolding for students to write their code around, and requires students to accurately implement the algorithms covered in lecture. The theoretical component of the HW assignments, which is due before the programming component, is designed to help the students understand the algorithms ``on paper'' before they are to implement them. The students also have weekly quizzes and two exams during the semester that test the concepts seen in class, and there is a final project where students work in teams of 2 or 3. We offer over 30 office hours per week and a Piazza board, and utilize undergraduate TAs due to their many benefits in the classroom (as discussed in \cite{Mirzaetal19}).

\subsection{Motivation for the Project}
Over the past several years, the programming assignments for this course have been graded by standard test-case based automatic grading programs along the lines of Autolab~\cite{Milojicic11}. These programs have streamlined grading dramatically and allowed for highly consistent grading; however, they do not have easy mechanisms to provide students with highly specific feedback. For example, the grading program for students' A* search implementation would fail an incorrect implementation, providing a starting state for which the implementation did not take the optimal path to the goal state, but it did not provide any explanation for which part of the student's code was causing the failure or award any partial credit.

To provide students with more helpful feedback, the instructors for the course considered two possible modifications to the course's grading: developing a program that gives students more granular feedback, and writing feedback by hand in order to give the most detail possible. To maintain consistency, manual graders would also base their feedback on the findings of a testing program.

It was resolved to trial both of these options in order to answer the following two questions: (1) do specific references to the syntax of a student's program help the student better understand the logic of the algorithm? and (2) does feedback on the style of a student's code help them write readable and maintainable code well-suited to collaboration? This experiment revealed evidence that (1) holds, but little evidence for (2). In addition, unexpectedly, students in the middle two quartiles of the human feedback group performed much better overall than those that received computer feedback.

\section{Methods}

After receiving IRB approval, we took the following approach: at the beginning of the semester, the class was randomly divided into two groups. The programming assignments of the first group were graded entirely by an in-house grading tool, and those of the second group were graded by teaching assistants with reference to the tool's output. A common difficulty facing researchers in education is the slew of biases that can be introduced when a new method is tested out during a semester. By splitting a single class into two groups, many variables such as instructor motivation or quality of materials were kept the same between the two groups, in a way they could not be over two semesters. In order to maintain fairness, the grade distributions of each programming assignment were closely compared and ensured to be virtually identical. Additionally, following the end of the semester, corrective action was taken to ensure that the course grade distributions of the two groups were comparable.

\subsection{Tool-based treatment}
Students in the tool-based grading group were graded by a grading tool developed specifically for this experiment. As opposed to test case-based automatic grading where edge cases for each criterion are devised and tested, this grading tool functioned by checking the calls made by student code for every logical element of a proper implementation of the algorithm in question. It makes heavy use of the flexibility of the Python interpreter, renaming functions and providing inputs designed to trace students' logic. This means that partial credit can be assigned according to students' adherence to the logic of the algorithm, rather than by the proportion of test cases they solved satisfactorily. One particular advantage of this approach is that code that never returns anything can still receive partial credit. Our grading tool is highly well-suited to an intermediate class in which students implement specific algorithms covered in lectures. See Figure 1 for a sample of the grading tool's output.
\begin{figure}
    \centering
    \includegraphics[width=\columnwidth]{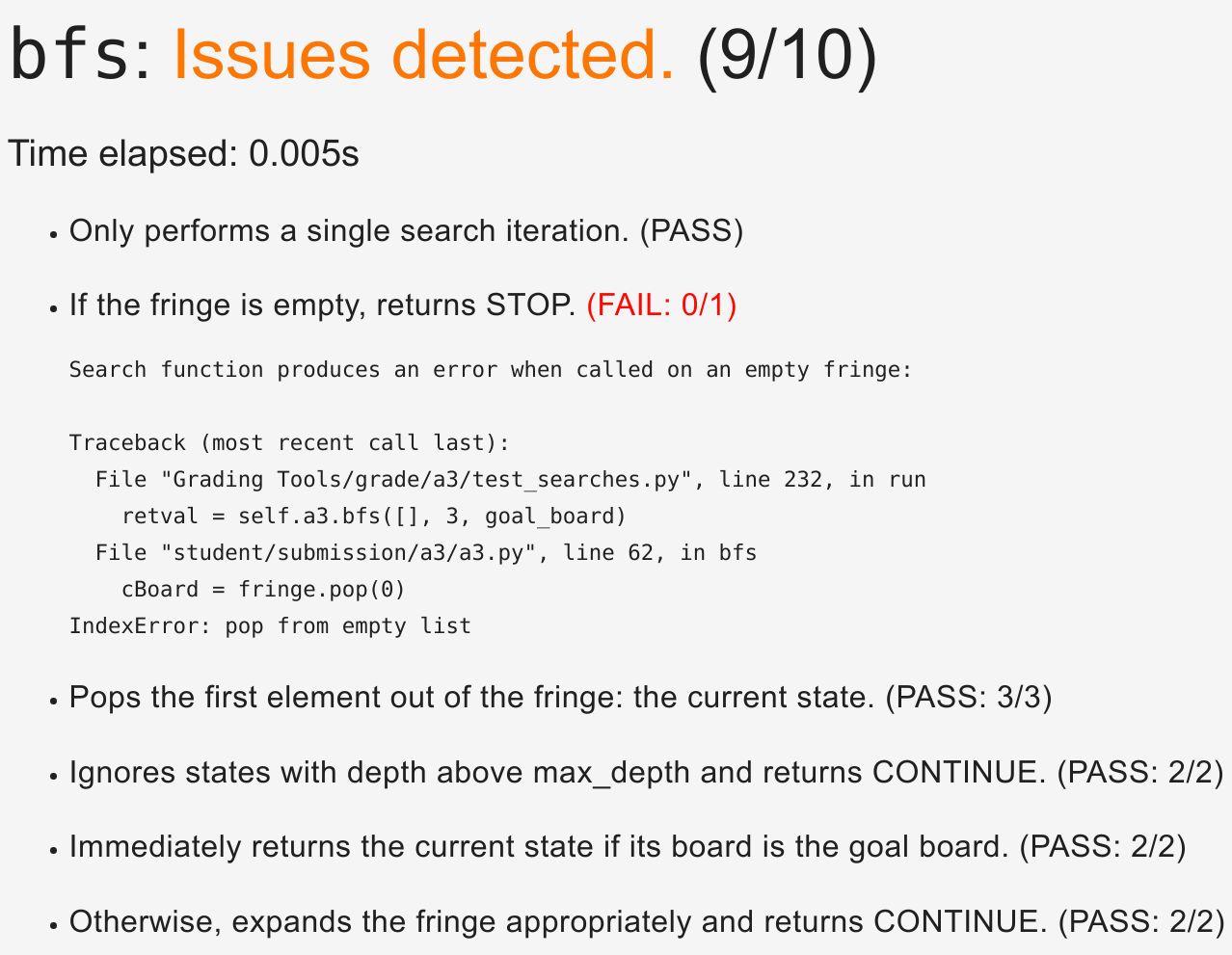}
    \caption{Sample output of the grading tool on an iteration of breadth-first search. Students in the automatic feedback group received only this logic feedback; those in the human feedback group also received specific references to syntax.}
    \label{fig:Figure1}
\end{figure}

\subsection{Human-assisted treatment}
Students in the human-assisted grading group generally received the output of the grading tool; however, human graders reviewed this output, adding clarifications when necessary, and added two additional types of feedback difficult for computers to generate.

First, graders added specific references to the syntax that caused students' code to fail a criterion -- a type of introspection that we cannot reliably perform in an automatic way. For example, this is the grader feedback for the student code from Figure 1:

\begin{displayquote}
Your condition ``fringe == None'' is really close to what you want. You need to know when the fringe is an empty list, so you can check if it's equal to the empty list, or you can check if it evaluates to False in a conditional.
\end{displayquote}

In addition, manually graded students received style feedback on the clarity, efficiency, and syntax usage of their code. Clarity feedback primarily focused on students' documentation, whitespace usage, and variable names. Efficiency feedback was largely based on time and memory complexity. Syntax feedback assessed students' adherence to Python's PEP8 style guidelines~\cite{PEP8} and the readability of their syntax choices. For example, students were encouraged to directly iterate over lists and generators wherever possible, rather than iterating over indices. A small number of points were allocated to these criteria to encourage students to take this feedback seriously. See below for some examples of grader style feedback:

\begin{displayquote}
In new\_heuristic, more descriptive variable names could be used for x and y.

Try not to use the length of a list when checking to see if it's empty.
\end{displayquote}

To ensure consistency, each student submission was evaluated by two graders, who rotated by assignment so that every hand-graded student would get feedback from the same distribution of graders over the course of the semester.

\subsection{Measures}
It was necessary to develop several measures to test the authors' hypothesis that students given specific references to the logic of their syntax would understand their code's function, and ultimately the algorithms they are implementing, better. A number of assessments, including quizzes and exam questions, were written to measure students' deeper understanding of the algorithms they implemented on the homework assignments. These assessments specifically focused on the logical steps of the algorithms implemented for the programming assignments, but were otherwise quite similar to the understanding-based questions on exams from previous semesters of the class. Additionally, one quiz was written to assess students' algorithmic thinking at the beginning of the semester, before they had received any feedback. All of these measures are available upon request.

To test our hypothesis that students given feedback about their code's style would write code better suited to collaboration, we analyzed the two groups' performance on the end-of-semester final project. This project has been used in the class for some time to help students develop their collaborative programming skills. In order to facilitate comparison of the two groups, project teams were chosen to each contain students from only one out of the two groups.

We excluded students who received failing grades in the class and students who dropped the class from our sample. Most of these students only made submissions for one or two of the assignments.

\section{Data}

\begin{table*}[h!]
\small
\begin{tabular}{|l||c |c |c|c|c|c|c|} 
 \hline
 & Quiz 1 & Quiz 5 & Quiz 8 &Exam 1.1 & Exam 1.2 & Exam 1.3 & Exam 2.7 \\ 
 & (10pts) & (10pts) & (10pts) & (10pts) & (10pts) & (10pts) & (13pts) \\ 
 \hline
 Mean (human-assisted) & 8.82 & 9.40 & 7.38 &6.64 & 7.68 & 4.94 & 11.58 \\
 \hline
 Mean (tool-based)    & 8.97 & 8.74 & 7.51 &5.57 & 7.2 & 3.17 & 12.14 \\ 
 \hline
 Significance value & 0.72 & 0.067 & 0.81 & 0.19 & 0.52 & 0.073 & 0.22 \\
 \hline
 95\% conf. interv. on diff. of means & (-0.99, 0.69) & (-0.07, 1.40) & (-1.23, 0.97) &(-0.56, 2.69) & (-1.03, 2.00) & (-0.16, 3.70) & (-1.47, 0.33) \\
 \hline
 Sig. values on mid 2 quartiles & 0.80 & 0.068 & 0.68 & 0.17 & 0.18 & 0.15 & 0.54 \\
 \hline
 95\% c.i. on mid 2 quartiles & (-1.64, 1.26) & (-0.05, 1.52) & (-1.32, 0.87) &(-0.72, 3.99) & (-0.56, 2.88) & (-0.63, 4.23) & (-2.08, 1.06) \\
 \hline
\end{tabular}
\caption{Performance on assessments of student understanding of algorithm logic}
\end{table*}

\begin{table*}[h!]
\small
\begin{tabular}{|l||c |c |c||c|c|c|c|c|c|} 
 \hline
 & \multicolumn{3}{|c||}{Final Project} & \multicolumn{6}{|c|}{Overall Class Data} \\
 \hline
 & Efficacy & Code Style & Report & Quizzes & HW & Exam 1 & Exam 2 & Office Hour & Final Grade \\ 
 & (20pts) & (10pts) & (25pts) & (100pts) & (100pts) & (100pts) & (100pts) & Visits & (100\%) \\ 
 \hline
 Mean (human-assisted) & 17.6 & 8.81 & 23.2 & 88.5 & 88.5 & 75.1 & 84.4 & 3.67 & 87.84 \\
 \hline
 Mean (tool-based)    & 17.6 & 8.69 & 22.1 & 88.5 & 88.6 & 71.8 & 80.2 & 6.20 & 86.31 \\ 
 \hline
 Significance value   & 0.97 & 0.64 & 0.26 & 0.98 & 0.97 & 0.35 & 0.27 & 0.16 & 0.46 \\
 \hline
 95\% conf. interv. on diff. & (-1.74, 1.81) & (-0.38, 0.62) & (-0.87, 3.02) & (-6.03, 6.14) & (-6.03, 5.82) & (-3.7, 10.3) & (-3.3, 11.7) & (-6.11, 1.04) & (-2.54, 5.60) \\
 \hline
 Sig. vals on mid 2 quartiles   & 0.28 & 0.10 & 0.40 & 0.97 & 0.91 & 0.16 & 0.040 & 0.20 & 0.010 \\
 \hline
 95\% c.i. on mid 2 quartiles & (-3.45, 0.97) & (-0.10, 1.33) & (-1.39, 3.42) & (-4.38, 4.50) & (-4.46, 5.00) & (-2.2, 13.5) & (0.9, 17.7) & (-10.6, 2.2) & (0.78, 4.46) \\
 \hline
\end{tabular}
\caption{Performance on collaborative final projects and overall class data}
\end{table*}

\begin{table*}[h!]
\small
\begin{tabular}{|l||c |c |c|c|c|c|c|c|c|c|c|c|c|c|c|} 
 \hline
Grade Distribution & Withdrew & F & D- & D & D+ & C- & C & C+ & B- & B & B+ & A- & A & A+ & Total\\ 
 \hline
 Human-assisted    & 8    & 3 & 0 & 1 & 1 & 1 & 2 & 2 & 2 & 3 & 3 & 7 & 9 & 2 & 44\\
 \hline
                   & 18\%   & 7\% & \multicolumn{3}{|c|}{5\%} & \multicolumn{3}{|c|}{11\%} & \multicolumn{3}{|c|}{18\%} &\multicolumn{3}{|c|}{41\%} & 100\%\\  
 \hline
 Tool-based        & 8    & 3 & 0 & 0 & 1 & 1 & 3 & 1 & 2 &12 & 5 & 4 & 4 & 2 & 46\\ 
 \hline
                   & 17\%   & 7\% & \multicolumn{3}{|c|}{2\%} & \multicolumn{3}{|c|}{11\%} & \multicolumn{3}{|c|}{41\%} &\multicolumn{3}{|c|}{22\%} & 100\%\\  
 \hline
 Total             & 16   & 6 & 0 & 1 & 2 & 2 & 5 & 3 & 4 &15 & 8 &11 &13 & 4 & 90\\ 
 \hline
                   & 18\%   & 7\% & \multicolumn{3}{|c|}{3\%} & \multicolumn{3}{|c|}{11\%} & \multicolumn{3}{|c|}{30\%} &\multicolumn{3}{|c|}{31\%} & 100\%\\ 
 \hline
 Previous Semester    & 22   & 1 & 0 & 2 & 2 & 2 & 7 & 9 &13 &10 &12 & 6 & 5 & 9 &100\\ 
 \hline
                   & 22\%   & 1\% & \multicolumn{3}{|c|}{4\%} & \multicolumn{3}{|c|}{18\%} & \multicolumn{3}{|c|}{35\%} & \multicolumn{3}{|c|}{20\%} &100\%\\ 
 \hline
\end{tabular}
\caption{Overall grade distributions}
\end{table*}

To determine to what extent the two groups can be statistically distinguished, Welch's T-test analyses were performed on all numerical data collected about students, including scores on problems, final project assessments, and overall class data. To avoid cherry-picking biases, only results on the problems and assignments that were written or selected for this project before the semester began are reported and discussed. Additional data are available upon request.

After seeing the grade distribution for the class we decided to additionally report this information because it is so striking. A very large percentage of students in the manual grading group received A grades in the class overall, whereas a wide plurality of students in the automatic grading group received Bs. At the end of this section, we have additionally included statistical analyses of the middle two quartiles of each group, where the overall grade distribution suggested the greatest effect. As it was conducted post-hoc, interpretation of the grade distribution and middle quartiles should be considered limited to qualitative discussion.

\subsection{Understanding of algorithm logic}
The results for Quiz 1, which was intended to ensure that the two groups were reasonably comparable, indeed suggest that the randomly selected groups of students have similar prior aptitudes. However, results on the later Quiz 5, designed to assess students' understanding of the backtracking algorithm, are highly suggestive ($p=0.067$) that students who received human feedback gained a better understanding than those who received only computer-generated feedback. Quiz 8, which asked students to extend their understanding of the alpha-beta pruning technique beyond that covered in class, does not show any significant improvement from the human feedback ($p=0.81$), though it is worth noting that the automatic feedback group received a very slightly higher mean score than the human assisted group.

To further test the students' understanding of the algorithms, we designed questions for the two exams  that would use the algorithms they had implemented in novel ways. For example, Exam 1 included a series of questions on which students applied the A* search algorithm (which was covered on a previous homework assignment) to an unfamiliar domain. We label the questions by Exam $X.Y$ in Table 1, where $X$ denotes the exam (1 or 2) and the $Y$ denotes the specific question in the exam. Concretely, the students were asked to perform a simple breadth-first search on a question in Exam 1.1, to construct a heuristic for the domain on question Exam 1.2, and to apply the heuristic as in the A* algorithm (calculating $f$-values and forming a priority queue) on Exam 1.3. Students who had received human feedback outperformed those who received computer-generated feedback on all three questions ($p=0.19$; $p=0.52$; $p=0.073$). Although manually-graded students outperformed computer-graded students on Exam 2 overall (Table 2, $p=0.27$), question Exam 2.7 was specifically written to assess conceptual understanding of decision trees, which were covered on the homework. Students were asked to apply a decision tree to various inputs, and then provide inputs that would minimize and maximize the execution time of recursively applying the tree. The results on this question actually suggested higher performance among students who had received automatic feedback ($p=0.22$).

\subsection{Style and collaboration}
Analysis on the grading rubrics for the class's final projects show that students from the human-assisted treatment received marginally better scores for coding style and on their final written reports, but these differences fall significantly below the threshold of statistical significance ($p=0.64$; $p=0.26$) and no difference whatsoever was found on the degree to which students accomplished their goals in the projects (efficacy).

\subsection{Overall performance}
Most of the differences in student grades occurred in the exam category. Although hand-graded students performed better on several specific quizzes, after allowing students to drop their lowest two quiz scores, the two groups' results were highly similar ($p=0.98$). So were their homework scores, which included both the written and programming assignments ($p=0.97$). It is on the examination scores that a macro-level difference is apparent: manually-graded students outperformed their automatically-graded peers by an average of 3.3 points on Exam 1 ($p=0.35$) and an average of 4.2 points on Exam 2 ($p=0.27$). It is also worth note that the average student receiving human feedback attended office hours 3.7 times over the semester, compared with 6.2 times for the average student receiving computer feedback ($p=0.16$).

When examining means, there is not a sizable difference between the two groups' final grades. However, the grade distributions reveal a stark difference: almost twice as large of a proportion of human-graded students received A grades compared with computer-graded students; conversely, more than twice as many B grades were awarded to students in the computer-graded group. This difference was not detected by mean/standard deviation techniques because the manually graded group has a heavier lower tail compared with the automatically graded group. However, the distribution is highly similar for all students who received below B's and also at the top of the A range, suggesting that the greatest effects of feedback are felt by students who are a priori more or less ordinary: neither the students who struggle the most, nor those who are prepared to excel without any intervention.

\subsection{Analysis on the middle quartiles}
Since the grade distribution reveals the greatest effect for ``average students'', we have additionally included statistical analyses conducted on middle two quartiles of each treatment group. Because we decided to conduct them after beginning analysis of our results, they must be considered strictly supplementary interpretation of our results, without the same statistical weight that accompanies direct testing and analysis of our hypotheses.

For the most part, analyses on the middle quartiles resemble those conducted on the full treatment groups; however, many of the results are less pronounced due to the loss of statistical power from halving group size. One notable exception is Exam 1.2, which yielded mixed results on the entire groups ($p=0.52$) but a clearer advantage for manually graded students when only the middle quartiles are considered ($p=0.18$). Another is the style component of the final project grades, where little effect could be discerned on the entire groups ($p=0.64$) but a more significant effect was revealed on the middle quartiles ($p=0.10$).

It is also worth noting that tool-graded students in the middle quartiles actually outperform their manually-graded peers by 2.2 points on efficacy ($p=0.28$), whereas the entire groups were indistinguishable on this criterion. This is likely caused by two outlier students in the manual group whose project received a score of 7.5 on efficacy. The lowest score any other student in the middle two quartiles received was a 14 out of 20.

Results are most striking on the exams ($p=0.16$, $p=0.040$) and on the final grade ($p=0.010$). (Markings of significance are omitted because of the post facto nature of this analysis.) Given that the middle-quartile students in each group performed very similarly on the initial standardizing quiz ($p=0.80$), these figures show that ``average'' students in the manually graded group had quantifiably better outcomes in the class than those in the automatically graded group.

\section{Discussion}\label{s:discussion}

\subsection{Understanding of algorithm logic}
Our results partially support the hypothesis that human grading helps students better understand the logic of the algorithms covered. In particular, manually graded students outperformed automatically graded students on Quiz 5, which reviewed the backtracking algorithm on the same Sudoku task covered in the programming assignment, and on Exam 1.1, 1.2, and 1.3, which applied the A* search algorithm to a novel task.

However, they did not even marginally outperform automatically graded students on Quiz 8, which asked students to analyze the alpha-beta pruning optimization in a way not discussed on the assignment. Neither did they outperform those students on Exam 2.7, where most of the points were an easy application of the recursive decision tree application algorithm, but the remaining points were dedicated to an unfamiliar run-time analysis of the algorithm.

All of these results hold for both the full groups and the middle-quartile students in each group.

One plausible interpretation of these mixed results is that feedback on code's syntax helps students understand the logic of their implemented algorithms better over the course of the semester, including when the algorithms are applied to novel tasks as on Exam 1.1, 1.2, and 1.3. However, this improvement would not extend to questions that ask students to analyze the algorithms in ways not covered in class. As this interpretation was not conceived prior to the study's execution, further research is needed to determine the extent to which it holds more generally.

An alternative interpretation of these results might be that after the first portion of the semester, students who had been receiving automatic feedback realized that they were not meeting their performance goals and spent more effort during the second half of the semester. However, the overall results of exam 2 do not support this interpretation: the performance gap was both wider and more significant on the second exam than on the first (see Table 2).

We conclude that receiving human feedback very likely improves students' conceptual understanding, but that this improvement is limited to certain, as-of-yet undetermined, areas. We posit that students who receive human feedback are better able to grasp the logic of the algorithms taught, but that this better grasp does not significantly improve their ability to perform higher-level analysis of that logic.

\subsection{Style and collaboration}
Most of our results suggest that human style feedback had little effect on students' final projects. When analysis is restricted to the middle two quartiles, several small effects are revealed: students who received human feedback appeared to do better on code style and worse on efficacy.

It is not difficult to imagine that giving students feedback on code style would result in them doing better when later graded on the same points of style. However, that it should result in them producing less successful projects is tougher to swallow. Efficacy was assessed as the degree to which final project groups met the goals they had set for themselves when proposing their final projects. Perhaps manually graded students' more promising grades earlier in the semester led them to propose more ambitious projects, but the difference in means is more likely due to an outlier effect from the two manually graded students who received far lower scores than any of the others.

All in all, we conclude that our results do not indicate that feedback on coding style improves the results on subsequent collaborative projects as originally hypothesized. There may be some extent to which students remember good style pointers, but even this is not clear and may not justify the extensive labor of giving students style feedback.

\subsection{Overall performance}
In spite of the previous equivocal results, our analyses of students' overall performance revealed that students who received human feedback had a (sub-significant) performance advantage on the exams over those who received automatic feedback, and that this advantage translated to higher overall course grades. This effect is particularly pronounced among those students who performed in the middle two quartiles of their treatment groups -- whom we call ``average students''.

In most courses, there will always be some students who are well-enough motivated and prepared that they will excel no matter what, and also students who are underprepared and need more support than the course staff is ready to offer. In this course, those students either did not need or could not receive detailed feedback from graders. One student who received an A in the class reported never opening a single feedback report -- they had already gotten full credit on all of the assignments. On the other hand, our graders often had difficulty providing feedback to struggling students whose submissions amounted to three or four lines of first approach. We therefore find it unsurprising that the most pronounced effects occurred for students in the middle two quartiles, and would strongly encourage further research to examine interventions for CS students in the bottom quartile.

As discussed previously, manually-graded students' higher performance likely stems in part from their greater facility with the logic of the algorithms taught in class, and likely not from their style feedback. However, we would also like to discuss two additional suggested mechanisms for this phenomenon. First, it might be proposed that the simple knowledge that graders would read students' code would provide a significant incentive for students to write more complete and polished code than they would otherwise, increasing the effort they put into assignments. Indeed, this effect would be greatest among average students, who are capable of writing very good code, but only with effort. However, the similar homework grades for the two groups, along with our evidence that the two groups were a priori very comparable, seem to indicate that the two groups spent more-or-less the same effort on their assignments.

Second, it might be proposed that receiving human feedback has a naturally demystifying effect: that a student might be unsure whether they are interpreting the grading tool's output correctly, and that human feedback clarifies any such ambiguities for students. This interpretation is in fact consistent with the algorithm logic mechanism discussed above. Clarifying ambiguities intuitively ought to lead to greater facility with concepts, because unambiguously understanding how one's code relates to a checklist of logical criteria helps one better understand how these criteria are put into practice. However, our graders' references to syntax go beyond mere disambiguation of error messages; they also directly connect even an unambiguously identified error in students' logic to the code that produced it. This deepens students' code-logic relation, and provides something even a perfect test case-based automatic grader cannot.

Finally, it is worth noting that human feedback appears to fulfill a similar need to office hours. Students who received human feedback visited office hours roughly half as often as those who received automatic feedback. This both suggests that disambiguation is one main purpose of human feedback, and suggests an exciting further direction. Massive Open Online Courses, an emerging trend in CS education, are generally automatically graded. It may, however, be worth dedicating some degree of instructor time to providing human feedback on student submissions to compensate for the impossibility of holding office hours. We strongly encourage extensions of our research to MOOCs.

\subsection{Challenges}
One constant challenge facing the instruction team was to ensure that students in the two cohorts were treated consistently. Human graders were able to understand when students were very close to satisfying criteria and award partial credit, and at the end of the semester we found that although programming assignment grades were overall very similar, automatically graded students at the bottom of the distribution for each assignment were receiving lower scores than equivalent manually graded students. After all assignments had been turned in, two of the TAs reviewed all of those students' submissions and awarded partial credit by the same criteria used for the manually graded students. Although this required significant labor on the part of the instruction team, it ensured that the grading on every assignment had been equivalent.

Indeed, labor was one of the biggest disadvantages of conducting this experiment. Even with 10 TAs, students received feedback on their homework assignments slightly later than promised multiple times throughout the semester.

Another issue observed was the use of the grading tool as a black-box debugger since we offered students unlimited opportunities to ``test their code'' for an estimated score but no feedback -- the precise opposite of the way students were intended to use it. One student tested their code on an assignment over 200 times.

\subsection{Future Semesters}
As a result of our observations and to target those students that will most benefit, we are switching from a group-based treatment to offering human feedback on request after students receive their grading tool results. This human feedback will be similarly detailed to the feedback on student logic provided in the study, and serves the same purpose of connecting students' logic with their code.

We are eliminating style feedback and grader-assigned partial credit. The style feedback appeared to have little effect on students' performance on their collaborative final projects, and the students who most often received grader-assigned partial credit were those in the bottom quartile, who did not exhibit significant improvement from receiving human feedback.

To deter the use of the grading tool as a black-box debugger, we are replacing the estimated-score test with a no-score compilation check, and extending the number of graded submissions from 2 to 15. In addition, we are awarding full credit to students' first three submissions. We intend this to encourage students to submit their work for feedback before it is perfect and to debug more mindfully. We hope this will result in students connecting their code with the algorithms' logic even without specific references to their syntax.

While our work strongly suggests that human feedback can lead to better student outcomes, it lacks statistical significance due to its small sample size. We strongly encourage fellow educators to pursue replications of this work, with a particular focus on specific references to syntax as a possible mechanism.

\section{Acknowledgements}
We thank the Spring 2019 CSCI-B 351 TAs, S. Dauer, P. Francis, L. Grim, G. Halloran, J. Mitchell, Z. Monroe, J. Nixon, and N. Pizzato. In particular, Kyle Yohler's web interface to our grading tool vastly simplified the process for both the students and the graders. 

\bibliographystyle{ACM-Reference-Format}
\bibliography{sigcse2020}

\end{document}